\documentclass{svjour3}                     % onecolumn (standard format)
\smartqed  % flush right qed marks, e.g. at end of proof
\usepackage{multirow}
\usepackage{url}

\usepackage{graphicx}

\usepackage{amssymb}
\usepackage{amsmath}
\journalname{Journal of Computational Electronics}
\begin{document}

\title{An efficient algorithm to calculate intrinsic thermoelectric parameters based on Landauer approach %
}

%\subtitle{Methodology and computational details}


\author{Abhijeet Paul         \and
        Shuaib Salamat \and
	Changwook Jeong \and
	Gerhard Klimeck \and
	Mark Lundstrom
}

\authorrunning{Paul et. al} % if too long for running head

\institute{   Abhijeet Paul, Shuaib Salamat, Changwook Jeong, Gerhard Klimeck and Mark Lundstrom\at
              School of Electrical and Computer Engineering and \\
              Network for Computational Nanotechnology \\
              Purdue University, West Lafayette, USA 47907 \\
              Tel.: 1-765-404-3589 \email{abhijeet.rama@gmail.com}           
}

\date{Received: date / Accepted: date}

%\begin{frontmatter}




%\author{Abhijeet Paul}
%\ead{abhijeet.rama@gmail.com}
%\author{Shuaib Salamat, Changwook Jeong, Gerhard Klimeck, Mark Lundstrom}

%\address{ School of Electrical and Computer Engineering and Network for Computational Nanotechnology,
% Purdue University, West Lafayette, IN, USA 47906 }

\maketitle

\begin{abstract}

The Landauer approach provides a conceptually simple way to calculate the intrinsic thermoelectric (TE) parameters of materials from the ballistic to the 
diffusive transport regime. This method relies on the calculation of the number of propagating modes and the scattering rate for each mode. The modes are calculated from the energy dispersion (E(k)) of the materials which require heavy computation and often supply energy relation on sparse momentum (k) grids. Here an efficient method to calculate the distribution of modes (DOM) from a given E(k) relationship is presented. The main features of this algorithm are, (i) its ability to work on sparse dispersion data, and (ii) creation of an energy grid for the DOM that is almost independent of the dispersion data therefore allowing for efficient and fast calculation of TE parameters. %The inclusion of scattering effects is also straight forward. 
The effect of k-grid sparsity on the compute time for DOM and on the sensitivity of the calculated TE results are provided. The algorithm calculates the TE parameters within 5\% accuracy when the K-grid sparsity is increased up to 60\% for all the dimensions (3D, 2D and 1D). The time taken for the DOM calculation is strongly influenced by the transverse K density (K perpendicular to transport direction) but is almost independent of the transport K density (along the transport direction). The DOM and TE results from the algorithm are bench-marked with, (i) analytical calculations for parabolic bands, and (ii) realistic electronic and phonon results for $Bi_{2}Te_{3}$.  

%The timing estimates for the calculations are also provided. The limitations of the algorithms are also pointed 
%out to make the users aware of the errors introduced in the calculations for sparse E(k) data.

%\PACS{}
\end{abstract}

% \linenumbers

\section{Introduction}
\label{intro}

\begin{figure}[!hbt]
\centering
\includegraphics[width=2.3in,height=1.4in]{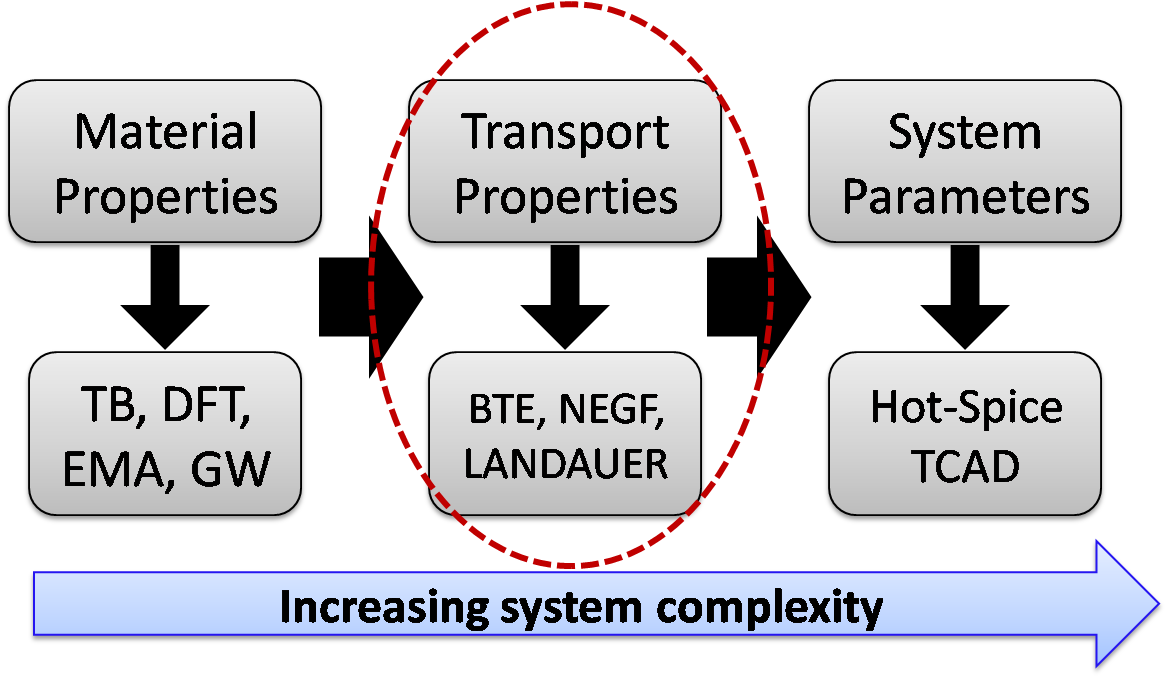}
\caption{Modeling hierarchy for the thermoelectric analysis. The present work focuses on the calculation of the TE parameters from the energy dispersion relations as shown by the encircled part. }
\label{fig:sim_hierarchy}
\end{figure}

Thermoelectricity provides an attractive and a clean way of converting waste heat into electricity. There have been a lot of efforts to improve the efficiency of thermoelectric (TE) devices. Solid-state TE devices are aggressively pursued both in the industry and research due to their advantages such as, (i) compactness, (ii) resistance to wear and tear, and (iii) portability. Thermoelectric efficiency ($ZT$) improvements need very careful engineering designs and optimization in terms of, (i) materials \cite{TE_mat_1,TE_mat_2,TE_mat_4,TE_mat_5}, (ii) structures like superlattices, nanocomposities, etc. \cite{TE_str_1,TE_str_2,TE_str_3,TE_str_4}, and (iii) devices \cite{TE_dev_1,TE_dev_2,TE_dev_3,TE_dev_4,TE_dev_5}. With so many design parameters it is extremely difficult to experimentally test every possible combination. At this point computer modeling plays a very significant role in designing and optimizing TE devices from material to the system level \cite{TE_model_1,TE_model_2,TE_model_3}. The present work focuses on the calculation of the TE transport parameters using the material energy dispersion as shown in Fig. \ref{fig:sim_hierarchy}.
%With more accurate bandstructure calculations available for electrons and phonons, an efficient calculation of modes will allow for the benchmarking and prediction of thermoelectric parameters for new materials. 

In the present work we focus on the calculation of the material properties involved in the calculation of $ZT$. The value of $ZT$ and the thermoelectric power-factor ($PF$) for a material are given by \cite{jeong_electron,kim_dim_TE,BTE_1}, 
\begin{eqnarray}
\label{eq_zt}
ZT & = & \frac{G \cdot S^2}{\kappa_{e}+\kappa_{l}} \cdot T \quad [unitless], \\
PF & = & G \cdot S^2 \quad [W/K^2 m^{d-1}] 
\end{eqnarray} 
where $G$, $S$, $\kappa_{e}$ and $\kappa_{l}$ are the electronic conductivity, electronic Seebeck coefficient, electronic thermal conductivity and lattice thermal conductivity, respectively. The term `$d$' is the dimensionality of the conductor. All the TE parameters depend on the electronic and lattice properties of the material.These material properties are strongly coupled and an improvement in one of the coefficients may degrade the other \cite{BTE_1}.

The Boltzmann transport equation (BTE) \cite{BTE_1,BTE_2} has been the most commonly used method to calculate the TE material parameters. However, with reduced dimensionality of the TE materials (like nanodots, nanowires, etc) the application of the Landauer approach \cite{Land,LAND_2} for calculating the TE parameters has gained a lot of attention \cite{jeong_electron,kim_dim_TE,jeong_phonon,TE_land_1,TE_land_2,mingo_ph,Mingo_kappa} due to the simplicity of the approach. The Landauer approach is applicable from the ballistic to the diffusive regime of transport for nanostructures. This model is insightful for understanding the impact of dimensionality on TE parameters too \cite{kim_dim_TE}. 

\begin{figure}[!hbt]
\centering
\includegraphics[width=2.3in,height=1.9in]{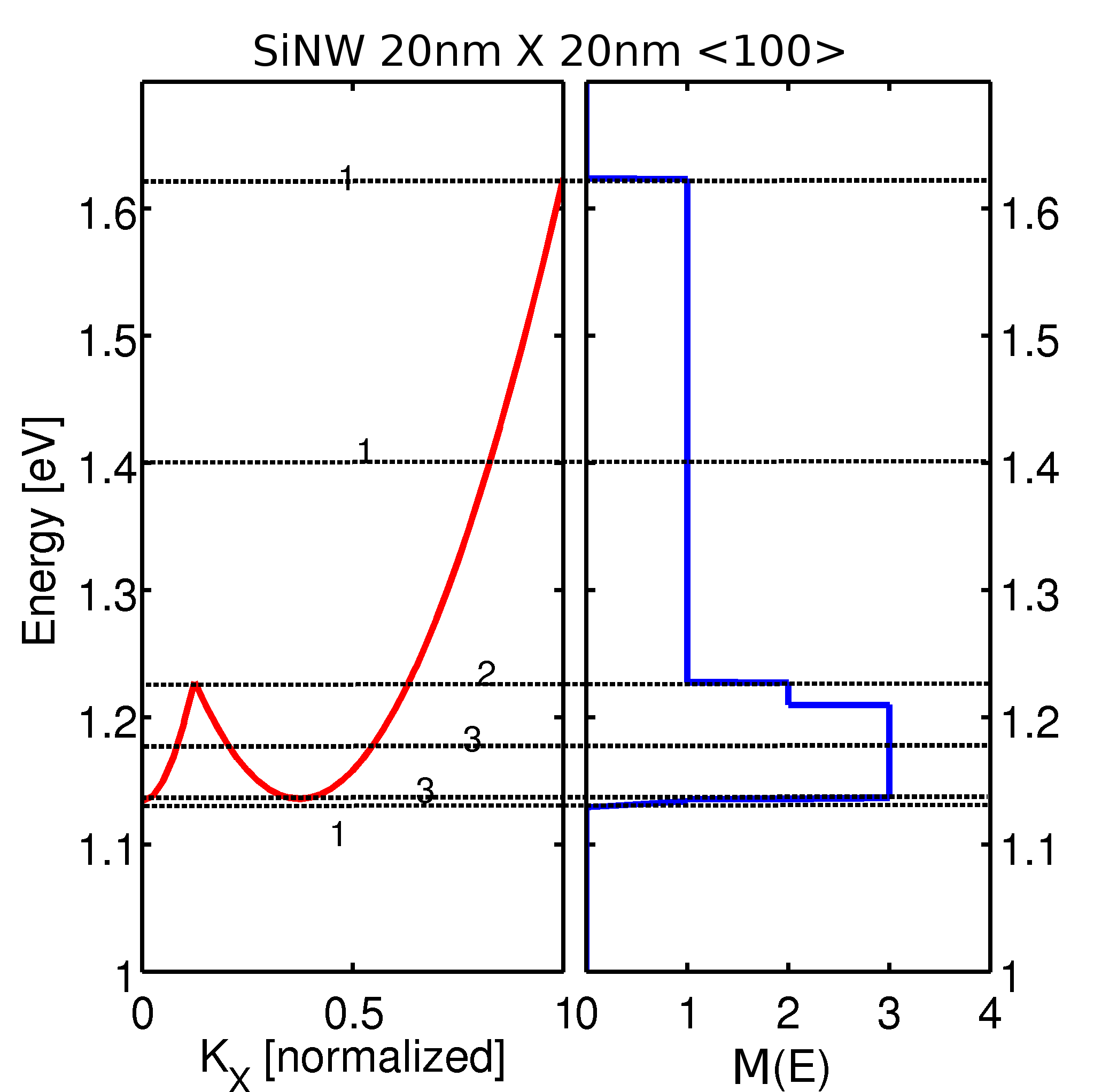}
\caption{The band-counting method for calculating the DOM. (a) The lowest sub-band of the electronic E(k) of a 20nm $\times$ 20nm [100] SiNW. (b) The corresponding propagating modes M(E) associated with this single band. }
\label{fig:sinw_modes_count}
\end{figure}

At the core of the Landauer approach is the calculation of distribution of modes (DOM) \cite{jeong_electron,jeong_phonon,kim_dim_TE} which is similar to the transport distribution function (TDF) used in the BTE model \cite{sofo_mahan} as shown in Refs. \cite{jeong_electron,jeong_phonon} for both electrons and phonons. The DOM represents the number of conducting channels available for the carriers, like electrons or phonons, at a given energy. From computational aspect, most of the previous work using Landauer's approach relied heavily on very fine E(k) calculations and then calculating the DOM by band-counting method \cite{jeong_electron} as shown in Fig. \ref{fig:sinw_modes_count}. The BTE methods use the reduced Brillouin Zone (BZ) integration schemes \cite{BZ_special_1,BZ_special_2} to calculate the TE parameters. However, these approaches too depend on a fine momentum mesh for numerical integrations. The computation of the dispersion relations in novel materials require significant computational resources and in general delivers results on momentum meshes that are not dense enough to derive a complete DOM or TDF.

To overcome the above mentioned computational challenges an efficient algorithm to calculate the DOM (used in the Landauer model) from a given E(k) is outlined in this work. The present method has two advantages over the previous band-counting methods, which are, (i) the energy dispersion (E(k)) can be relatively sparse, and (ii) the energy grid for the DOM and the E(k) does not have to be identical. Overall compute time for the calculation of TE parameters is reduced in two steps, (i) relatively little compute time is needed to calculate the DOM from the sparse energy dispersion, and (ii) the sparse DOM energy grid further reduces calculation time of the TE parameters ($G$, $S$, $\kappa_{e}$, and $\kappa_{l}$).

%The Landauer approach is elegant and is extremely helpful as it allows to separate transport into Modes - only E(k) dependent quantity and Transmission–an E(k) and scattering dependent. The usefulness and advantages of Landauer approach have been discussed by in Refs.\cite{jeong_electron,jeong_phonon,TE_land_1,TE_land_2}. 

The present work is divided in the following sections. The basic TE theory in the linear transport regime is outlined in Sec. \ref{sec_1_1}. The generic algorithm for DOM calculation is presented in Sec. \ref{sec_DOM} with specific changes required for electrons in Sec. \ref{sec:ele_trans_kernel}, and for phonons in Sec. \ref{sec:phon_trans_kernel}. Discussion on the transmission calculation is provided in Sec. \ref{sec_Trans}. The results section provides the k dependent sensitivity analysis in Sec. \ref{sec:sensitivity} and the timing analysis of the algorithm in Sec. \ref{sec:timing}. Comparison and verification of the TE parameters, calculated using the algorithm, with published results are provided in Sec. \ref{sec:verification}. The summary of the work is outlined in Sec. \ref{sec:conc}.

%The method is applicable and extended to electron as well as phonons. The algorithm allows easy energy integration involving various scattering mechanism such as constant mean free path (mfp), constant scattering time (tau), energy dependent mfp and energy dependent tau. The paper is organized as follows. In section 2, we will discuss the theory and methodology. We also elaborate how the methodology differs from other works and how the algorithm works with sparse grid. In section 3, the method to compute thermoelectric parameters is presented. In section 4, test cases and verification results are presented.

\section{Theory and methodology}
\label{sec_1}

In this section the calculation of the TE parameters and the details of the algorithms are outlined.

\subsection{Thermoelectric parameters in the linear transport regime}
\label{sec_1_1}

The $ZT$ of a material at a temperature, $T$, is based on the calculation of the intrinsic material properties which include both the electronic and the lattice properties (see Eq. (\ref{eq_zt})). The electronic transport parameters are obtained using the Landauer approach in the zero current limit \cite{jeong_electron,jeong_phonon} as,

\begin{eqnarray}
\label{eq_G}
G & = & \frac{2q^2}{h} \cdot I_0 \quad [\Omega^{-1}m^{d-1} ] \\
\label{eq_S}
S & = & -[k_B/q]\cdot[I_1/I_0]\quad [V/K] \\
\label{eq_ke}
\kappa_e & = &[\frac{(2Tk_B^2)}{h}]\cdot[I_2-(I_1^2/I_0)]\quad [W/m^{d-1}K] \\
\label{electron_int}
I_j &=& \int_{E_{min}}^{E_{max}}\Big[\frac{(E-E_F)}{k_BT}\Big]^j \cdot \mathcal{T}(E) \cdot \mathcal{M}(E)\cdot \frac{-\partial \mathcal{F}_{FD}}{\partial E} \cdot dE,
\end{eqnarray}

where $I_{j}$ is the jth order energy moment integration around the Fermi Level ($E_{F}$). The terms $q$, $k_{B}$ and $h$ are the electronic charge, Boltzmann constant, and Planck's constant, respectively. In the quantity $I_{j}$ the terms $\mathcal{M}$(E), $\mathcal{T}$(E), and $\mathcal{F}_{FD}$ are the distribution of modes (DOM) at energy $E$, transmission at energy $E$ and the Fermi-Dirac distribution function, respectively.

The lattice thermal conductivity ($\kappa_{l}$) can be calculated using the Landauer's model as \cite{jeong_phonon,mingo_ph,Mingo_kappa},

\begin{eqnarray}
\label{ktherm_eq}
\kappa_{l}(T)& = & \hbar P_{1} \quad [W/m^{d-1}K] \\
\label{phonon_int}
P_{j} & = & \int_{\omega_{min}}^{\omega_{max}}\mathcal{T}(\omega)\cdot \mathcal{M}(\omega)\cdot\omega^{j}\cdot \\ \nonumber 
      &   &\frac{\partial}{\partial T} \Big[ (\exp(\frac{\hbar \omega}{k_{B}T})-1)^{-1}\Big]\cdot d\omega 
\end{eqnarray}

where $P_{j}$ is the jth order phonon energy integration. The terms $\mathcal{M}(\omega)$ and $\mathcal{T}(\omega)$ are the distribution of modes, and the transmission of the modes at a phonon frequency $\omega$, respectively.

Equations (\ref{electron_int}) and (\ref{ktherm_eq}) show that the calculation of any transport parameter within the Landauer model depend on two quantities, (i) the distribution of modes ($\mathcal{M}$) and, (ii) the transmission of the modes ($\mathcal{T}$). The DOM depends only on the energy dispersion of the carriers in the material whereas the transmission is controlled by the dispersion and the scattering mechanisms of the carrier. The advantage of the Landauer model lies in the separation of the transport kernel into two parts that can be solved using parallel computer programming leading to a faster and efficient calculation of the transport parameters. In the next part the details of the algorithm to efficiently calculate the DOM from a given energy dispersion is outlined.

%in 0D (dots), 1D (nanowires), 2D (thin films) or 3D (bulk) periodic material.
%The physical equivalence of LA and BTE have been shown by Changwook $et$ $al.$ \cite{jeong_electron}. 

%Under Landauer's approach (LA) the total transport kernel gets separated into 2 parts, (i) the density of modes (DOM) (M(E)) which depends only on the bandstructure of the material and, (ii) the carrier MFP ($\lambda(E)$) which depends both on the bandstructure and the scattering mechanisms present in the system which is absent in Boltzmann transport equation (BTE). 
\begin{table}[t!]
\centering
\caption{Dimensionality of structure and dependence on `K' vectors}%
\label{table_dim_k}
\begin{tabular}{|l|c|c|c|c|l|}
\hline
Structure & Periodic & Confined & $K_{\perp}$ & $ K_{\parallel}$ & K \\
(dimension) & dim. (P) & dim. (C) & & & \\
\hline
%Dots (0D) & 0 & 3 & 0 & 0 & $\vert$K$\vert$ = 0 \\
Wires (1D) & 1 & 2 & 0 & 1 &[$K_{\parallel}$]\\
Films (2D) & 2 & 1 & 1& 1 &[$K_{\perp},K_{\parallel}$]\\
Bulk (3D) & 3 & 0 & 2 & 1 &[$K^{1}_{\perp},K^{2}_{\perp},K_{\parallel}$]\\
\hline
\end{tabular}
\end{table}

\subsection{Distribution of Modes calculation}
\label{sec_DOM}

\begin{figure}[!hbt]
\centering
\includegraphics[width=3.1in,height=2.2in]{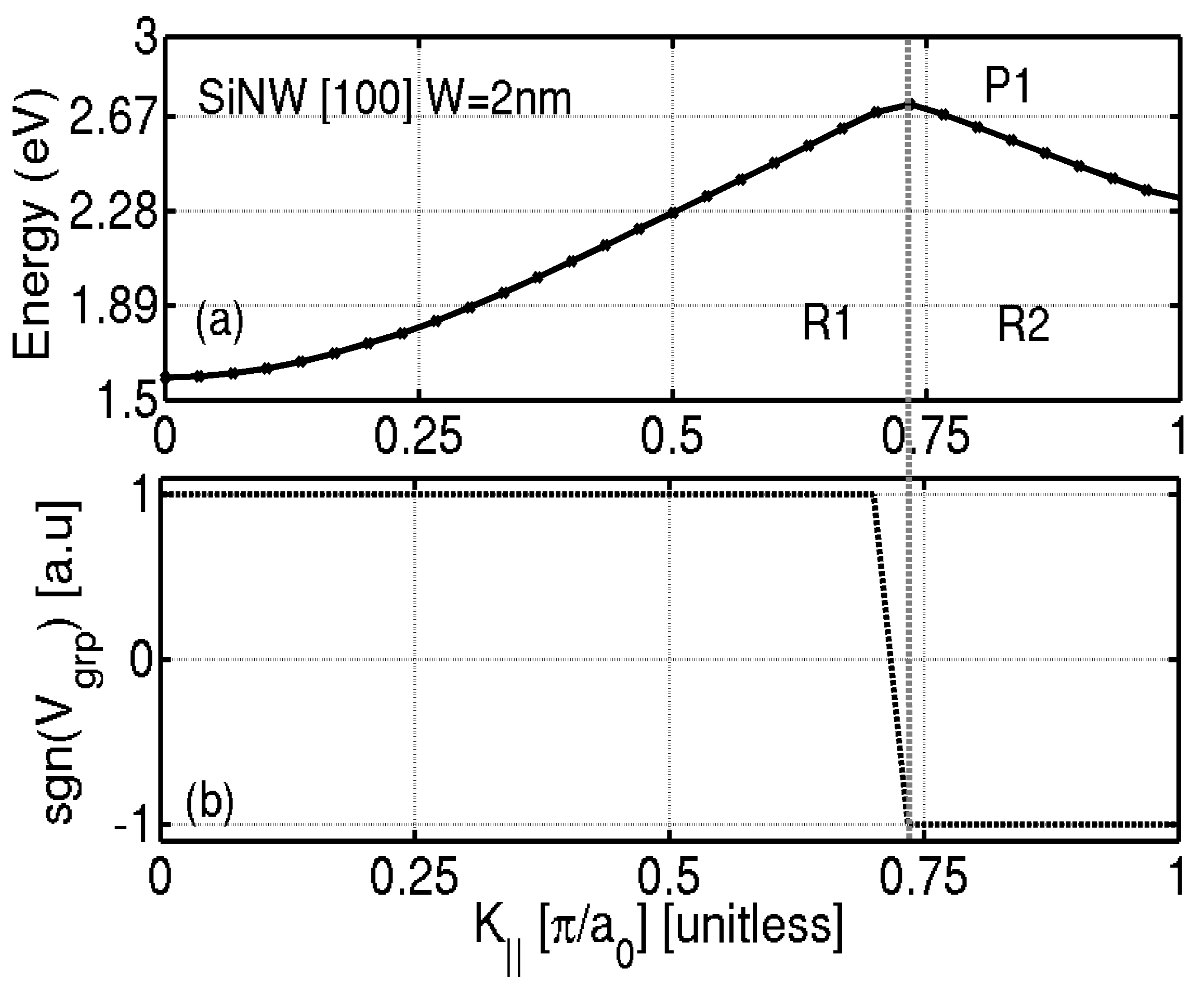}
\caption{Velocity at each point positive half of first conduction band sub-band for a 2nm X 2nm [100] Si nanowire (SiNW).
As mentioned above, the important points to note in the given band are the points where sign of velocity changes. These points are indicated as P1, P2, and P3 and corresponding regions of interest are marked as R1, R2, and R3 in Figure 2.3. Each point provides monotonic velocity range (increasing or decreasing) and calculations for DOM are performed separately for these ranges within a sub-band. }
\label{fig:vel_grid}
\end{figure}

\begin{figure}[!hbt]
\centering
\includegraphics[width=2.4in,height=2.4in]{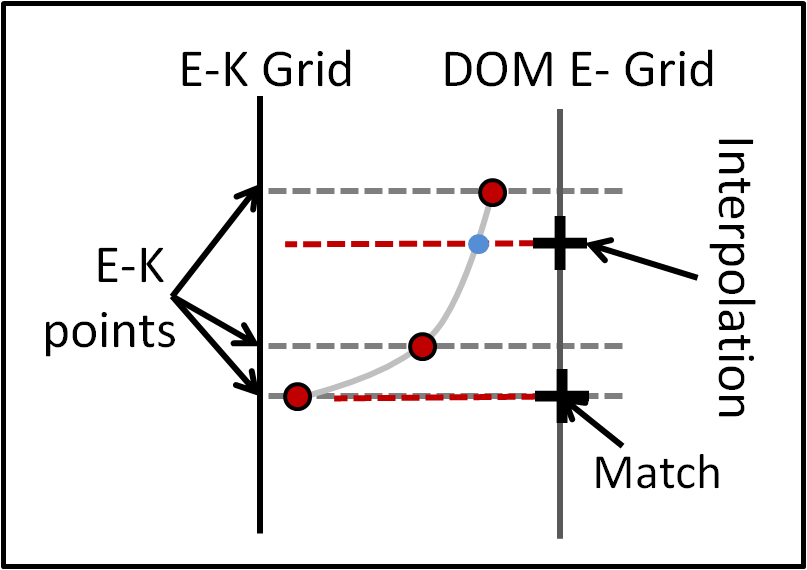}
\caption{The $E-K$ points on the provided energy dispersion shown by dots. The energy grid for DOM (EGD) points are shown using crosses. These points are of two types, (i) the matching point to the $E-K$ grid, and (ii) the one which requires interpolation of the provided $E-K$ relationship. This interpolation (either linear or quadratic) is done in the appropriate monotonic energy region like $R_{1},\;R_{2}$,etc shown in Fig. \ref{fig:vel_grid}. In this way the DOM is created for the EGD.}
\label{fig:interp_dom}
\end{figure}

The step by step procedure for the calculation of DOM (applicable to both electrons and phonons) is given below,

\begin{enumerate}
\item{Obtain the energy dispersion of an $d$ dimensional, where $d$ = 1, 2 or 3, periodic material. The momentum vector $'K'$ can be decomposed into two components, (i) along the transport direction denoted by $K_{\parallel}$, and (ii) in the direction perpendicular to the transport direction denoted by $K_{\perp}$ depending on the dimensionality of the conductor as shown in Table \ref{table_dim_k}.}

\item{For each combination of $K_{\perp}$, a 1D $E-K_{\parallel}$ is obtained which is used for mode counting. The energy grid for the DOM (EGD) is created based on the 1D $E-K_{\parallel}$ for all the $K_{\perp}$. This energy grid does not have to be identical to the energy values from the $E(k)$ data. The details of choosing the energy limits for the electrons and phonons are outlined in Sec.\ref{sec:ele_trans_kernel} and Sec. \ref{sec:phon_trans_kernel}, respectively. The energy grid is chosen so as to provide a reasonable compromise between the computation time and the accuracy of the results. }

\item{For a 1D $E-K_{\parallel}$, the group velocity ($v_{grp}$) is calculated to find out the regions of monotonic variation in the energy with $K_{\parallel}$. Only positive $K_{\parallel}$ are considered since the $E-K_{\parallel}$ relations are symmetric. The +ve half group velocity is calculated as, 

\begin{equation}
\label{eq:vgrp}
 v_{grp} = \frac{1}{\hbar} \frac{\partial E(K)}{\partial K_{\parallel}}	
\end{equation}

The monotonic velocity regions, R1 and R2, for an example $E(k)$ are shown in Fig. \ref{fig:vel_grid}. }

\item{ These monotonic velocity regions are then used for counting modes. The EGD points are calculated by the interpolation of the $E-K_{\parallel}$ data points using the $V_{grp}$. The details for calculating the energy nodes on the EGD is shown in Fig. \ref{fig:interp_dom}. As a by-product of the calculation the carrier velocity is also obtained which can be used for other calculations such as the mean free path. }

\item{The modes from each of the 1D $E-K_{\parallel}$ are then integrated over all the $K_{\perp}$ and divided by the unit-cell area ($A_{uc}$) to obtain the total DOM.}
\end{enumerate}

%The limits of the energy grid are defined within the minimum and cutoff of the given E(k) relationships. 
%Next step is to calculate velocity at each point on the given E(K). 
%These calculations are performed individually on each band and within a band at each K-point. 
%This process allows us to obtain regions were the velocity changes signs. Velocity is taken as slope of E(k)
%Therefore, focusing on the +k states, Fig. 2.1 shows the first subband and Fig. 2.2 shows the velocities calculated at each point on this subband.

\begin{figure}[!hbt]
\centering
\includegraphics[width=3.1in,height=2.2in]{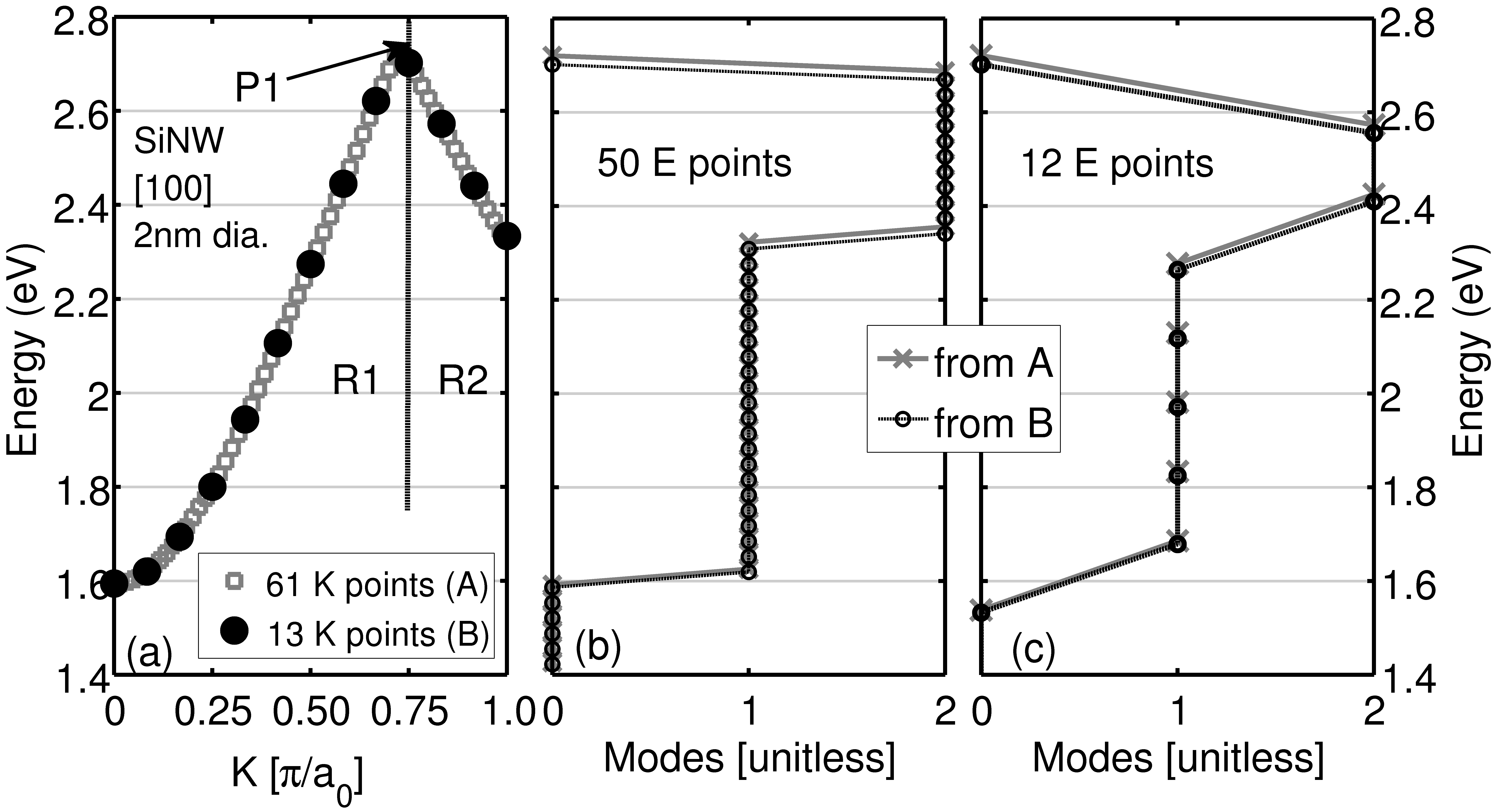}
\caption{(a) $E(K)$ relation with different number of k points. Case A with 61 K points and case B with 13 K points. Also the two monotonic $E-K$ regions are shown (R1 and R2) along with the turn around point P1. DOM calculated for the two $E-K$ grids using (b) 50 energy points and (c) 12  energy points. The DOM matches 100\% for all the 4 cases showing the robustness of the DOM calculation method. As long as the sparse $E-K$ captures the important turn around points (like P1) correctly the DOM calculation algorithm obtains the correct number of modes.}
\label{fig:add_dom}
\end{figure}

The present algorithm has the advantage that the original $E-K_{\parallel}$ can be sparse compared to the energy grid on which the DOM is calculated since the monotonic regions of $E-K_{\parallel}$ allow to interpolate the dispersion data to be used for DOM and velocity calculations. The strength of the algorithm to obtain the DOM for different K-grid densities is shown in Fig. \ref{fig:add_dom}. Since the algorithm can work on sparse dispersion data, the time to obtain the total modes is also reduced. 

The present algorithm assumes that all the $K$ vectors for a given energy dispersion are orthogonal. This assumption has both advantages and disadvantages. Since the $K_{\parallel}$ and $K_{\perp}$ are separated, this allows for parallel computation of modes for each $K_{\perp}$ set. This leads to computational speed-up. This aspect of the algorithm can be inspected as a future work. The interpolation in the E(k) is always done along $K_{\parallel}$ but not along $K_{\perp}$. This puts a limitation on the sparsity of the $K_{\perp}$ grid. A very sparse $K_{\perp}$ grid will result in erroneous DOM calculation. The over-all E(k) should not be too sparse either such that the original features of the dispersion are lost. In that case velocity interpolation will give erroneous results. The sensitivity of the TE results on the K-grid sparsity and the compute time for DOM are discussed in Sec. \ref{sec:sensitivity} and Sec \ref{sec:timing}, respectively.

Apart from the general steps adopted for the calculation of the DOM for both electrons and phonons, some special care in selecting the energy ranges for both, involved in Eq. (\ref{electron_int}) and (\ref{phonon_int}), must be taken.
%% Next the details about the implementation of these equations are provided.

\subsection{Energy range: electron transport}
\label{sec:ele_trans_kernel}

\begin{figure}[!htb]
\centering
\includegraphics[width=2.4in,height=1.7in]{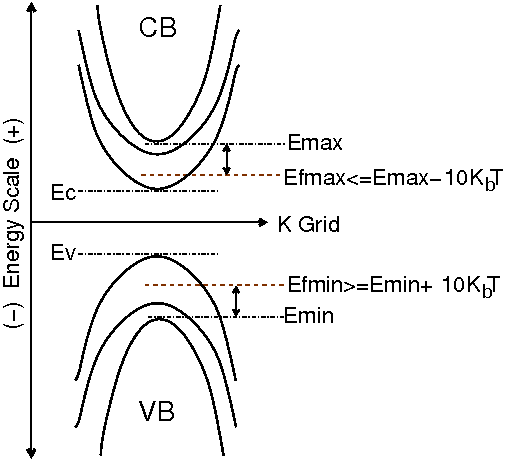}
\caption{Schematic showing the range of energy limit and the range of Fermi-level used for the calculation of the integral in Eq. \ref{electron_int} for electrons.}
\label{fig:Efs_range}
\end{figure}

\begin{figure}[!bht]
\centering
\includegraphics[width=3.4in,height=1.8in]{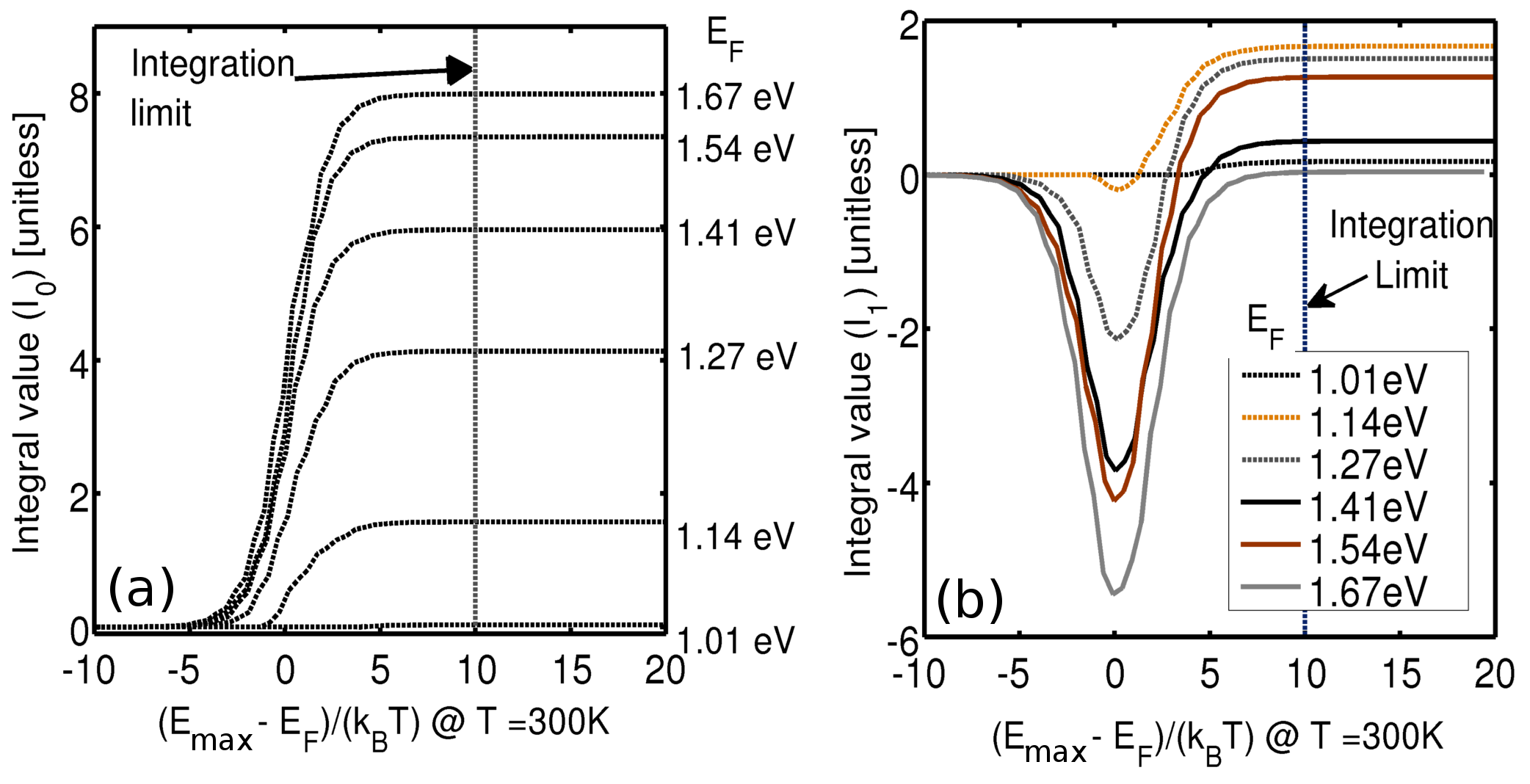}
\caption{Variation in the values of (a) $I_{0}$ and (b) $I_{1}$ (Eq. \ref{electron_int}) for different values of the Fermi-level ($E_{F}$). When the $E_{F}$ is at least 10$k_{B}$T below the $E_{max}$ then the integral values show less than 1\% variation. Similar result is also obtained for the integral $I_{2}$.}
\label{fig:integral_val}
\end{figure}

Real materials are characterized by many different electronic bands.  However, not all these bands contribute to the electron transport and an energy range around the Fermi-level ($E_{F}$) needs to be selected carefully.  To obtain an expedient but good approximate solution the energy cut-offs ($E_{min},E_{max}$) are chosen such that the integral values for the transport parameters ( Eq. (\ref{eq_G})-(\ref{electron_int})) do not show any variation. The bounds for the energy grid (Eq. \ref{electron_int}) of the DOM is obtained as follows, 

\begin{eqnarray}
\label{eq:ene_lim}
E_{max} &=& Ec \quad or \quad min[max[E(K_{\parallel}) \forall K_{\perp}]] \\
E_{min} &= & Ev\quad or \quad max[min[E(K_{\parallel}) \forall K_{\perp}]],
\end{eqnarray}

where $min (max)$ represent the minimum (maximum) value in a given numerical array. The terms $Ec$ and $Ev$ define the conduction band minima (CBM) and the valence band maxima (VBM), respectively as shown in Fig. \ref{fig:Efs_range}. Our calculations show that in order to obtain correct results, the $E_{F}$ value can vary between the following limit, 

\begin{equation}
\label{eq:ef_lim}
E_{min} + 10k_{B}T \le E_{F} \le E_{max} - 10k_{B}T
\end{equation}
where $T$ is the temperature. The choice of 10$k_{B}T$ is chosen since this gives a good range where the integrals involved in the TE parameter calculations become invariant to the choice of energy grid as shown in Fig. \ref{fig:integral_val}.

\subsection{Energy range: lattice transport}
\label{sec:phon_trans_kernel}

The lattice kernel calculations do not depend on any kind of Fermi-level. Unlike the electron bands, the phonon bands are always within a fixed energy range with a varying number of sub-bands for different dimensional structures \cite{jce_own_paper,iwce_own_paper}. Also there is no negative phonon energy dispersion in stable semi-conductor structures \cite{SINW_110_phonon}, hence the energy grid of the DOM always contains positive values. The energy limit for the lattice kernel is chosen as follows, 
%from 0 to the maximum phonon energy (or some user defined value in the allowed energy range).

\begin{eqnarray}
\label{eq:phon_ene_lim}
\omega_{min} &=& 0 \quad or \quad \text{User defined}  \\
\omega_{max} &= & \Omega_{max} \quad or \quad \text {User defined},
\end{eqnarray}

where $\Omega_{max}$ is the maximum energy limit of the phonon dispersion.

%The procedure works excellently even with sparse energy grid,to calculate thermoelectric parameters  as the band counting is not performed on the given dispersion relation. Rather the features of the given E(K) are captured by determining the monotonic velocity ranges and the actual mode counting is performed on the user define energy grid. This makes the algorithm fast and accurate. This does not mean that the k-mesh could be reduced arbitrarily. As we will show in Section 4, there are limitations to the reduction of k-points. Below some k-mesh the results for transport parameters start to depict large variations.

\subsection{Transmission calculation} 
\label{sec_Trans}
For ballistic transport of electrons or phonons the transmission ($\mathcal{T}$(E)) of all the modes is 1. However, in reality carriers undergo a lot of scattering which depends on the dimensionality of the system, doping, temperature, etc. This reduces the transmission of the modes below 1. For a conductor of length $L$, $\mathcal{T}(E)$ is given by \cite{jeong_electron,jeong_phonon}, 

\begin{equation}
\label{TE_def1}
	\mathcal{T}(E) = \frac{<\lambda(E)>}{L+<\lambda(E)>}	
\end{equation}

where $<\lambda(E)>$ is the carrier mean free path (MFP) obtained by the summing over all allowed $k$ points at energy $E$. When $L$ $>>$ MFP (diffusive limit) then Eq.(\ref{TE_def1}) can be approximated as, 
\begin{equation}
\label{TE_def2}
	\mathcal{T}(E) \approx \frac{<\lambda(E)>}{L}	
\end{equation}

All the scattering mechanisms present in a system are lumped in the `mean free path'. The energy dependence of the MFP can be broadly classified into two categories, (i) constant MFP (no energy dependence), and (ii) energy dependent MFP. For some common scattering mechanisms like ionized impurity, acoustic phonon, etc, $<\lambda(E)>$ can be expressed in a power law form as $<\lambda(E)>=\lambda_0 [E/(k_B T)]^r$, where $E$ is the kinetic energy of the carrier, $\lambda_0$ is a constant and $`r'$ is a characteristic exponent describing a specific scattering process \cite{jeong_electron}.

In most of the BTE calculations the scattering time ($\tau_{scat}$) is used instead of the MFP. Again for $\tau_{scat}$ the energy dependence are of two types, (i) energy independent (constant $\tau_{scat}$), and (ii) energy dependent. The constant $\tau_{scat}$ case is physically hard to justify since it is well known that particles scatter to/from different energy states at a different rate \cite{jeong_electron}. The connection of the scattering time to the MFP is given as \cite{jeong_electron},

\begin{equation}
\label{tau_to_lam}
	<\lambda(E)> = 2\cdot\frac{\sum_{K} v_{\parallel}^{2}(K,E)\cdot \tau_{scat}(K,E)}{\sum_{K}\vert v_{\parallel} (K,E) \vert}
\end{equation} 

Here the summation is over all the $K$ states at a given energy $E$. If the scattering time is assumed isotropic in $K$ then the MFP is given as, 

\begin{equation}
\label{tau_to_lam_iso}
	<\lambda(E)> = 2 \cdot \Big[\frac{\sum_{K} v_{\parallel}^{2}(K,E)}{\sum_{K} \vert v_{\parallel} (K,E) \vert} \Big] \cdot \tau_{scat}(E)
\end{equation} 
 
In the present algorithm, the MFP can be calculated efficiently since the velocities are already obtained as a by-product during the DOM calculation. In the case of a constant scattering time, the energy dependence of the MFP is derived purely from the electronic or phonon energy dispersion.

\section{Results and Discussion}
\label{sec:result}
 
In this section we provide the results on the dependence of calculated TE parameters on the sparsity of the K-grid using the algorithm. The timing analysis is provided to give an idea about the total speed up with K-grid reduction and which part of the calculation consumes the maximum compute time. Also the comparison of the DOM and TE parameters, calculated from the algorithm, with analytical expressions and realistic material dispersions are provided.

\subsection{Sensitivity analysis: How robust is the algorithm?}
\label{sec:sensitivity}

To understand the strengths and limitations of the algorithm, K-grid sensitivity tests are performed using parabolic E(k)
 dispersions for 3D, 2D and 1D cases. The parameters used for the generation of  the parabolic bands are shown in Table \ref{tab_eff_param}.

The TE parameters like G and S are calculated using the parabolic E(k). The k-grid density variation introduces numerical error in the S and G calculation. The percentage error in the maximum power factor ($PF_{max}$) is related to the error in $S$ and $G$ by the following relation,
\begin{equation}
\label{PF_err}
\frac{\Delta PF_{max}}{PF_{max}} = 2\cdot \frac{\Delta S}{S} + \frac{\Delta G}{G},
\end{equation}

where $\Delta$S/G are the variations in Seebeck coefficient and electronic conductivity, respectively. The final fluctuation in the $PF_{max}$ depends on the sign of $\Delta$S/G.  However, all the fluctuation plots are shown for the absolute value of the errors.
% Whenever the final error in $PF_{max}$ is smaller than the corresponding error in $S$ or $G$ indicates a sign change in either of the quantities.

To start the sensitivity analysis, first a base K-grid is chosen. A K-grid with 100 points in each direction (-$\pi/a_{0}$ to $\pi/a_{0}$) is found to be sufficient to provide stable results. A finer K-grid does not change the final results  by more than 0.5\% for any of the calculated TE values. Three different types of studies were performed to determine the sensitivity of the algorithm to reduction in $K_{\parallel}$ and $K_{\perp}$,

\begin{itemize}
	\item{Case A: Keep $K_{\perp}$ fixed at 100 grid points and reduce $K_{\parallel}$ down to $\sim$ 60\%-80\%.}
	\item{Case B: Keep $K_{\parallel}$ fixed at 100 grid points and reduce $K_{\perp}$ down to $\sim$ 60\%-80\%.}
	\item{Case C: Reduce both $K_{\parallel}$ and $K_{\perp}$ down to $\sim$ 60\%-80\%.}
\end{itemize}

\begin{table}[t!]
\centering
\caption{Parameters used for the generation of parabolic bands}%
\label{tab_eff_param}
\begin{tabular}{|l|c|c|c|c|c|c|}
\hline
Structure & $m^{*}_{\parallel}$ & $m^{*}_{\perp1}$ & $m^{*}_{\perp2}$ & $E_{c}$ & $E_{v}$& $a_{0}$ \\
(Dim) & $\times m_{0}$ & $\times m_{0}$ & $\times m_{0}$ & eV &  eV  & nm\\
\hline
Wires (1D) & 1 & -- & -- & 0.2 & -0.2 & 1\\
Films (2D) & 1 & 1 & -- &  0.2 & -0.2 & 1\\
Bulk  (3D) & 1 & 1 & 1 & 0.2 & -0.2 & 1\\
\hline
\end{tabular}
\end{table}

For the sake of brevity only the 2D TE error analysis results are shown. Other dimensions (3D and 1D) show similar results and the outcomes are similar. For the 1D system only case C is applicable since in these systems only $K_{\parallel}$ is the free momentum direction. The other two directions are geometrically confined as shown in Table \ref{table_dim_k}.

\subsubsection*{Sensitivity Analysis: Case A} 
\begin{figure}[!hbt]
\centering
\includegraphics[width=3.1in,height=1.8in]{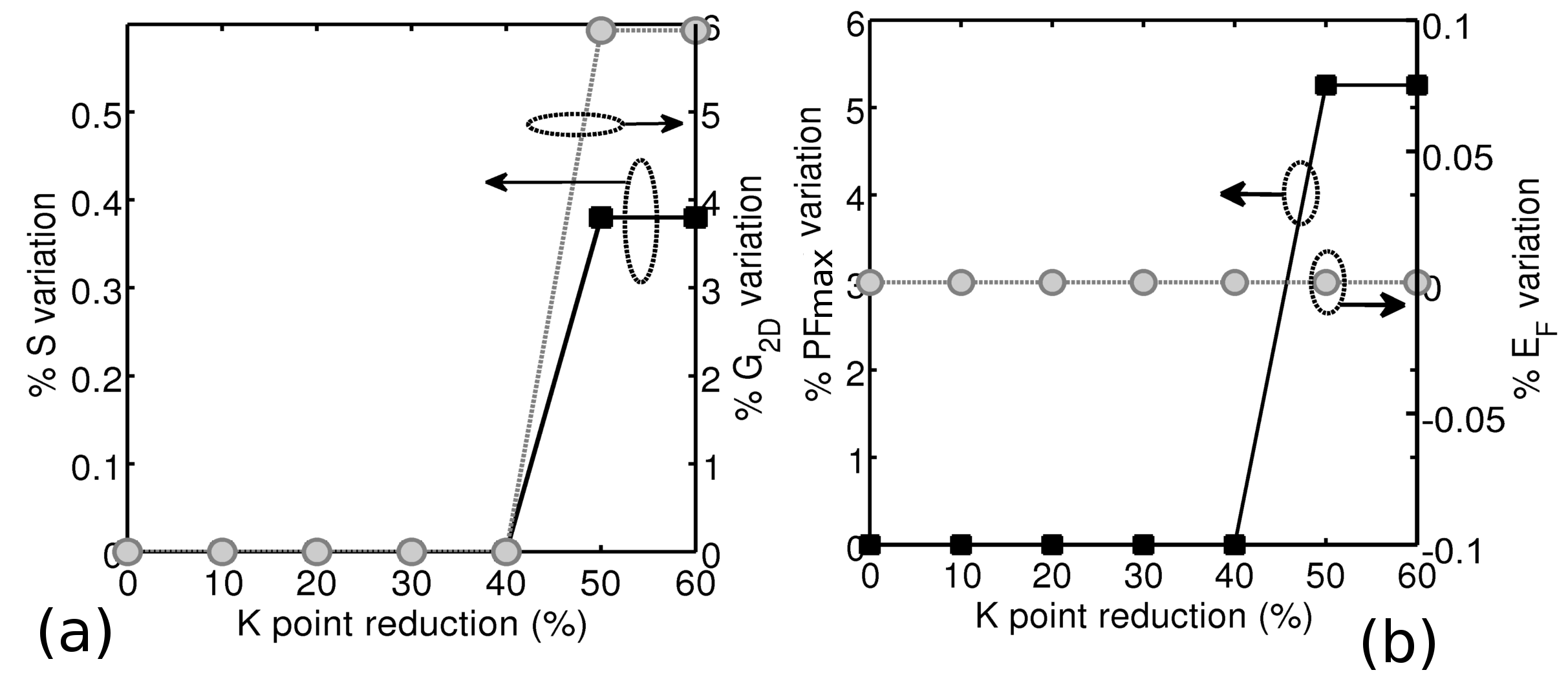}
\caption{Impact of $K_{\parallel}$ point reduction  on (a) $S$ (left) and $G$ (right) and (b) $PF$ (left) and $E_{F}$ (right) in a 2D structure. All the values are extracted at the maximum PF point. The $K_{\perp}$ has 100 grid points. Even for 60\% reduction in $K_{\parallel}$ points none of the TE values show more than 6\% variation.}
\label{fig:tpt_k_red_2D}
\end{figure}

The reduction in $K_{\parallel}$ down to 60\% results in less than 1\% variation in $S$ and $\sim$6\% variation in $G$ as shown in Fig. \ref{fig:tpt_k_red_2D}a. The corresponding fluctuation in the $PF$ is around 5\% as shown in Fig. \ref{fig:tpt_k_red_2D}b. The Fermi-level, at which the maximum $PF$ is extracted, however remains unchanged. The fluctuation in the TE parameters arises only from the fluctuation in the DOM. Thus, the present DOM calculation method is quite robust to reductions in $K_{\parallel}$ given the $K_{\perp}$ has good mesh density.

\subsubsection*{Sensitivity Analysis: Case B}
\begin{figure}[!htb]
\centering
\includegraphics[width=3.2in,height=2.0in]{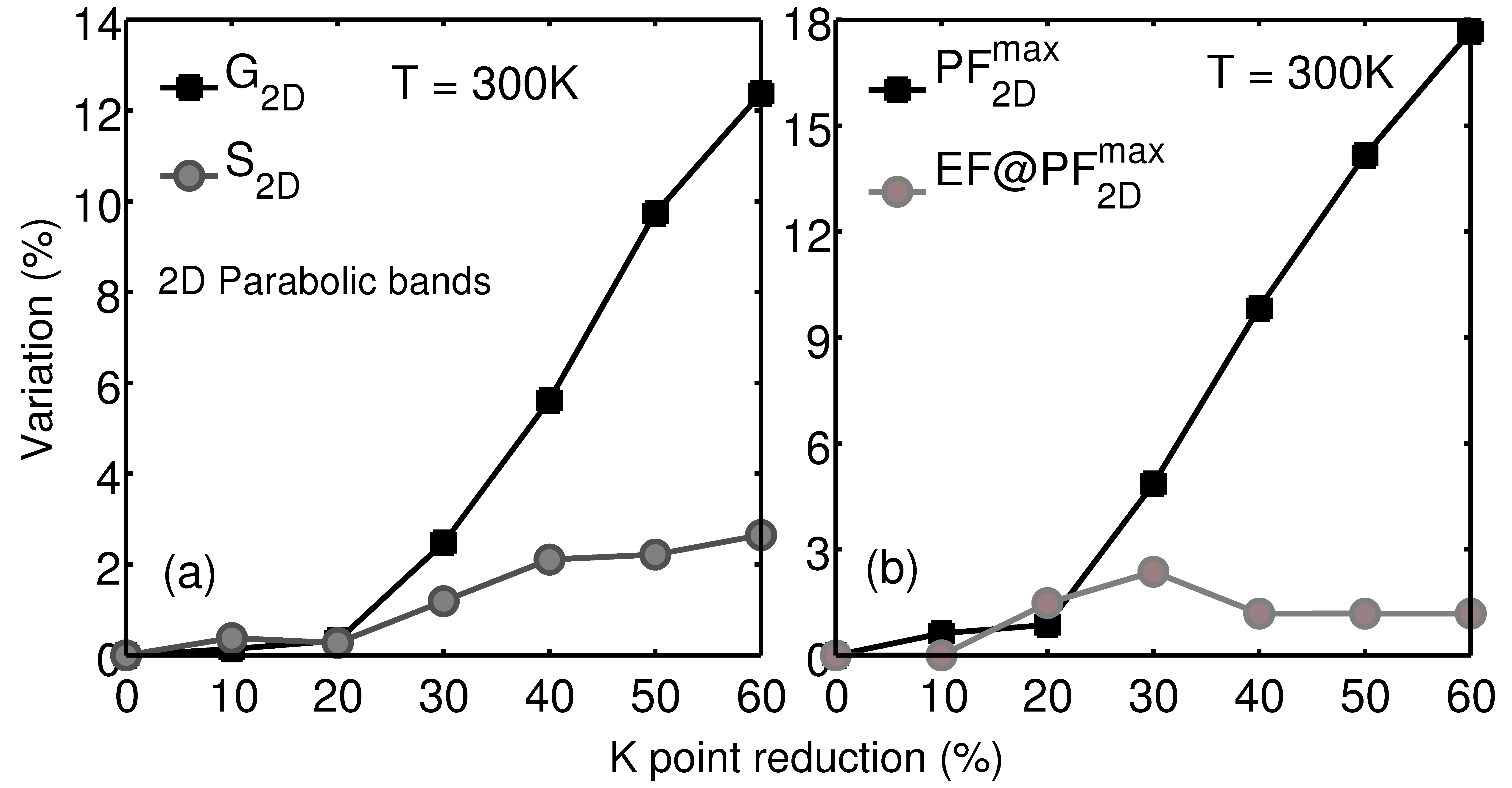}
\caption{Impact of $K_{\perp}$ point reduction  on (a) $S$ (left) and $G$ (right) and (b) $PF$ (left) and $E_{F}$ (right) in a 2D structure. All the values are extracted at the maximum PF point. The $K_{\parallel}$ has 100 grid points. For 60\% reduction in K-points $G$ shows a maximum variation of 12\% and $PF_{max}$ has variation around 10\%.}
\label{fig:norm_k_red_2D}
\end{figure}

The reduction in $K_{\perp}$ down to 60\% results in less than 2\% variation in $S$ and $\sim$12\% variation in $G$ as shown in Fig. \ref{fig:norm_k_red_2D}a. The maximum fluctuation in PF is around 10\% as shown in Fig. \ref{fig:norm_k_red_2D}b. The Fermi-level ($E_{F}$) at which the maximum $PF$ is extracted shows a maximum variation of $\sim$2.5\%. In this case, the fluctuation in the TE parameters arises from the fluctuation in, (i) the DOM, and (ii) the $E_{F}$. The present DOM algorithm is sensitive to variations in $K_{\perp}$. 

\subsubsection*{Sensitivity Analysis: Case C}
\begin{figure}[!t]
\centering
\includegraphics[width=3.2in,height=2.0in]{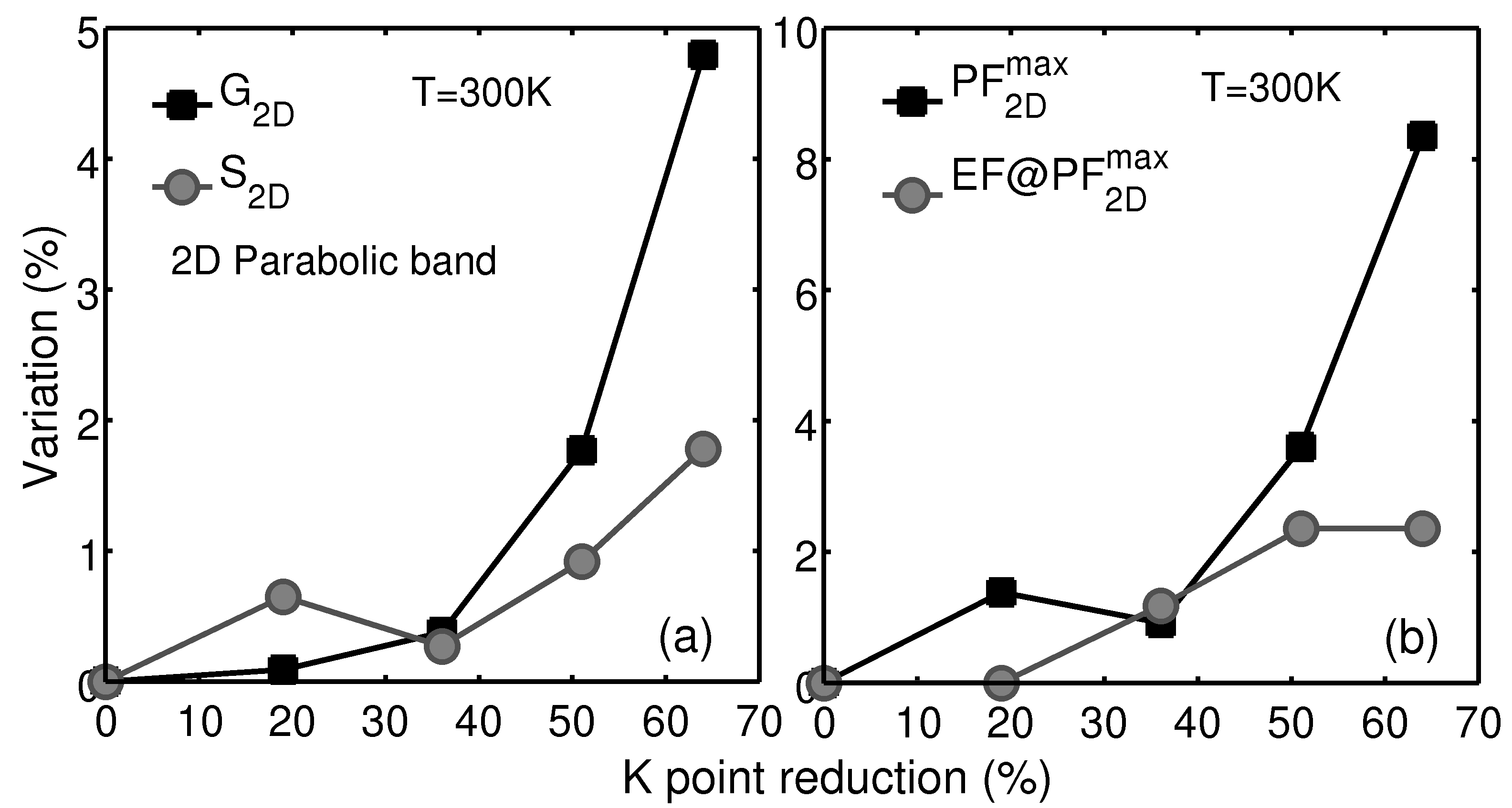}
\caption{Impact of reduction in all the K-points,  on (a) $S$ (left) and $G$ (right) and (b) $PF$ (left) and $E{F}$ (right) in a 2D structure. All the values are extracted at the maximum PF point. $G$ shows a larger fluctuation ($\ge$10\%) compared to $S$ fluctuation ($\le$4\%) which also reflects in the $PF_{max}$ fluctuation.}
\label{fig:all_k_red_2D}
\end{figure}

The reduction in all the $K$ points down to 60\% results in less than 5\% variation in $S$ and $\sim$13\% variation in $G$ as shown in Fig. \ref{fig:all_k_red_2D}a. The maximum fluctuation in $PF$ is around 10\% as shown in Fig. \ref{fig:all_k_red_2D}b. The $E_{F}$ at which the maximum $PF$ is extracted shows a maximum variation of $\sim$2.5\%. Thus, the fluctuation in the TE values arises from the fluctuation in, (i) the DOM, and (ii) the $E_{F}$. This case has almost similar K-grid sensitivity as case B, again showing that the present DOM algorithm is sensitive to variations in $K_{\perp}$.

\subsection{Timing analysis}
\label{sec:timing}

The present algorithm can calculate the TE parameters within reasonable error limits as shown in the previous section. Another obvious question that arises is how much computational speed-up can be achieved. The time to calculate the DOM for the three cases presented in the previous section is analyzed for 3D, 2D and 1D structures.  

\begin{figure}[!hbt]
\centering
\includegraphics[width=3.0in,height=2.6in]{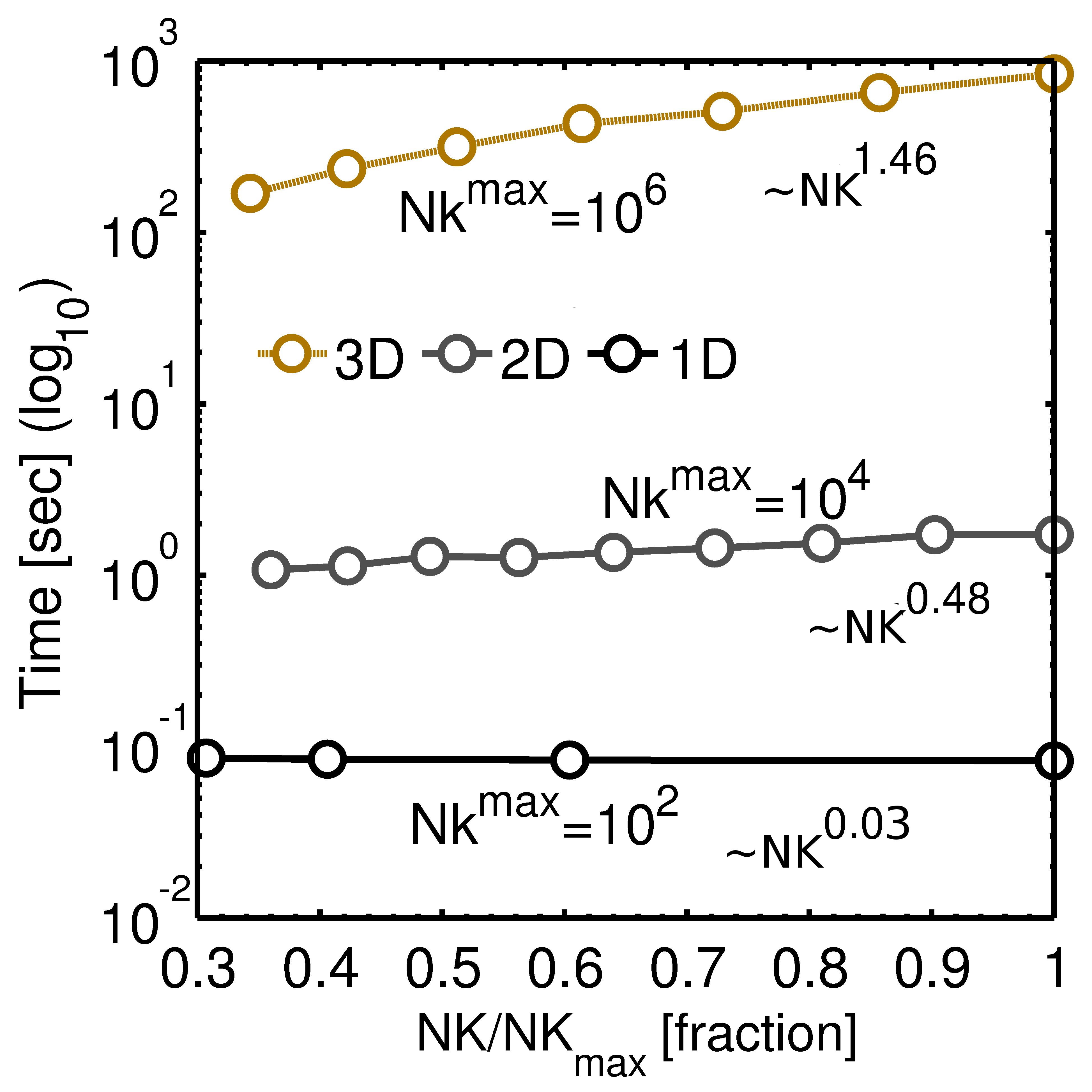}
\caption{DOM compute time for 1D, 2D and 3D parabolic bands. The number of K points are reduced along all the K-directions equally. The 3D case takes the maximum time due to higher number of K-points, followed by the 2D and the 1D case.}
\label{time_all_dim}
\end{figure}

As the $K$ density along all the directions is reduced, the speed up for each dimension is different. For the 3D system, the time required goes up with total number of K-points ($NK$) with a power of 1.46 ($NK^{1.46}$). For the 2D case the power law is $NK^{0.48}$ and for 1D case the time taken is almost constant (in the given $NK$ range). The algorithm takes roughly 900 seconds for 1 million K-points (100$\times$100$\times$100) for 3D case on nanoHUB.org workspace \cite{workspace}. For the 2D case the time taken for ten thousand K-points (100$\times$100) is nearly 2 seconds and for 1D case the time taken is roughly 0.1 second for 100 K-points. All these results are shown in Fig. \ref{time_all_dim}.

\begin{figure}[!htb]
\centering
\includegraphics[width=3.4in,height=1.8in]{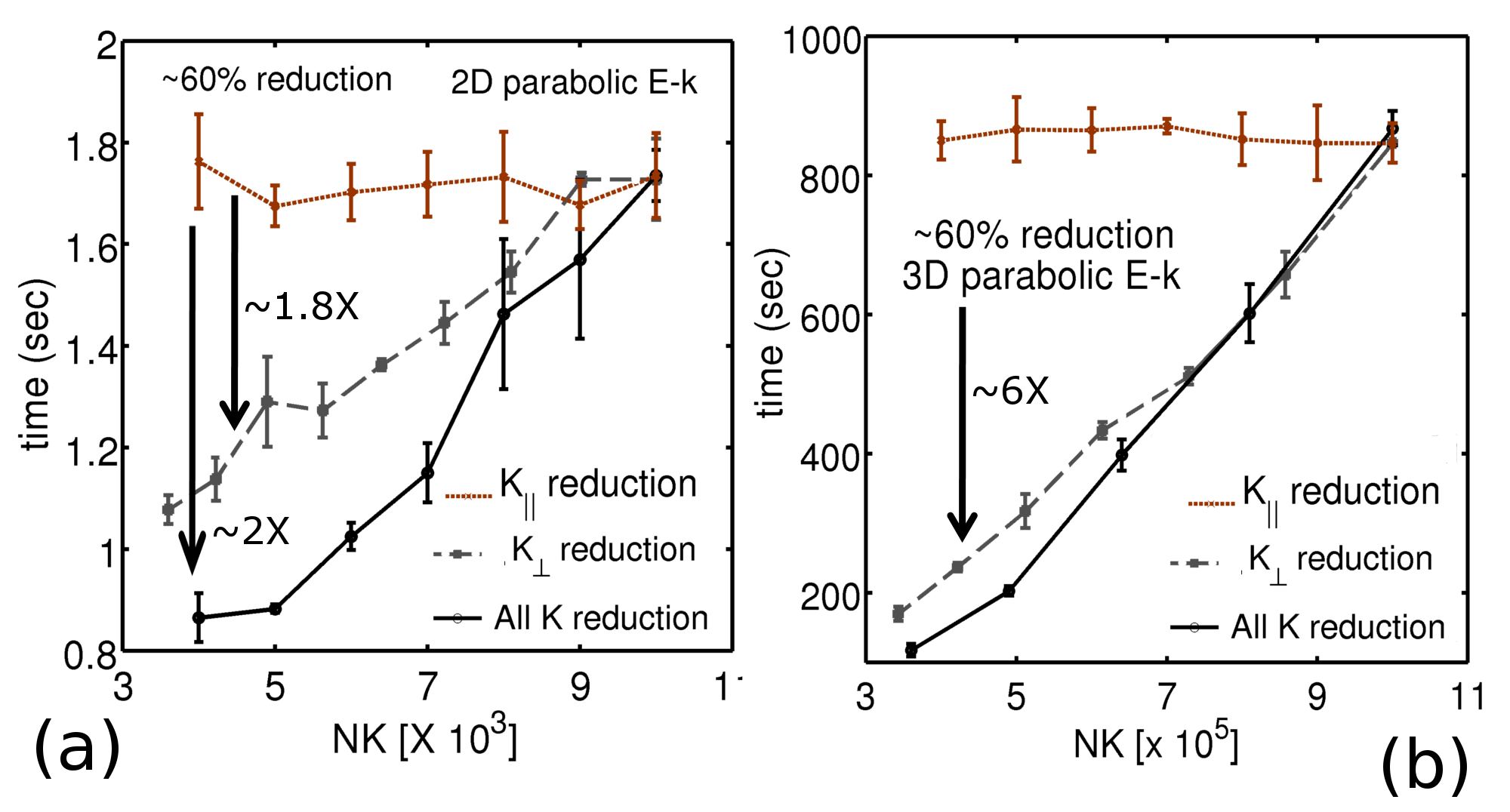}
\caption{DOM compute time ($t_{DOM}$) for the three types of K-point reduction for the (a) 2D structure and (b) 3D structure. For both the cases the compute time is almost independent of $K_{\parallel}$ reduction (brown line).  reduces  with $K_{\perp}$ point reduction. For each point 5 compute times are averaged. }
\label{time_all_2_3_dim}
\end{figure}

For the cases of 2D and 3D, the algorithm requires different compute times with reductions in K-points along both the transport and the transverse direction.  The compute time for the DOM ($t_{DOM}$) is roughly independent of the K-point reduction in transport direction for both the 2D and 3D cases (Fig. \ref{time_all_2_3_dim}). However, for a 60\% reduction in $K_{\perp}$, the 2D case shows a $t_{DOM}$ speed up factor of $\sim$2 (Fig. \ref{time_all_2_3_dim} a). While for the 3D case, a speed up factor of 6 is observed (Fig. \ref{time_all_2_3_dim} b). A reduction in all K-points along all directions show a similar speed up (Fig. \ref{time_all_2_3_dim}). Thus, the present algorithm shows a good speed up with $K_{\perp}$ point reduction. 

\subsection{Discussion: Algorithm aspects}

The TE sensitivity analysis and $t_{DOM}$ speed up reveal that the algorithm to calculate the DOM is more sensitive to the $K_{\perp}$ points compared to the $K_{\parallel}$ points. A reasonable reduction in $K_{\perp}$ must be chosen in order to optimize the compute time and to obtain reasonably stable TE parameter values. A summary of all the analysis is provided in Table \ref{table:tab_2}. This table also provides the limits for reduction in K points in the E(k) data-set to obtain TE parameters within a 10\% error margin. In most of the cases a 50\% reduction in K-points is easily achievable without a big penalty on the calculated TE values. The sensitivity analysis presented here is for parabolic bands,  however, the general features of the algorithm remain quite similar even for the dispersions of real materials which are more complex than parabolic bands. Similar conclusions are obtained for the phonon dispersions.

\begin{table}[!t]
\centering
\caption{Summary of sensitivity and timing analysis}%
\label{table:tab_2}
\begin{tabular}{|c||ccccccc||c|}

\hline
Dimension & K-grid   & Max. K  & \multicolumn{4}{c}{Max. error (\%)}& Good $E(K)$ & DOM speed up\\
	      & direction & reduction (\%) & S & G & PF & $E_{F}$ & sparsity  & (60\% K red.)\\
%age) Max sigma variance Max PF variance Max Ef variance Time reduction for reducing k by 60% (%age)
\hline
3D & $K_{\parallel}$ & 80 & -4.41 & 3.02 & -5.65 & 4.14 & $<$70\% & 1.02$\times$ \\ \hline
3D & $K_{\perp}$ & 64 & -5.3 & -4.45 & -11.49 & 1.02 & $<$50\% & 8.4$\times$ \\ \hline
3D &  All-K & 66 & -5.3 & 15.2 & 5.9 & 5.2 & $<$50\% & 3.6$\times$\\\hline \hline

2D & $K_{\parallel}$ & 60 & -0.4 & 6 & 5 & 0 & $<$60\%& 1.1$\times$\\ \hline
2D & $K_{\perp}$ & 60 & -0.32 & -9.30 & -9.95 & 1.18 & $<$50\%& 1.8$\times$ \\ \hline
2D & All-K & 58 & 4.2 & -13.3 & -4.4 & 1.18 & $<$50\% & 1.5$\times$ \\\hline \hline

1D & All-K & 80 & 4 & 13.2 & 20 & 3.8 & $<$70\% & 1.05$\times$ \\ \hline

\end{tabular}
\end{table}

\subsection{Calculation of the TE parameters}
\label{sec:verification}

The final verification of the algorithm is done by calculating the TE parameters for (i) the parabolic bands in 3D, 2D, and 1D cases, and (ii) bulk $Bi_2Te_3$. 

\subsubsection*{Parabolic Bands}

\begin{figure}[!hbt]
\centering
\includegraphics[width=2.1in,height=5.0in]{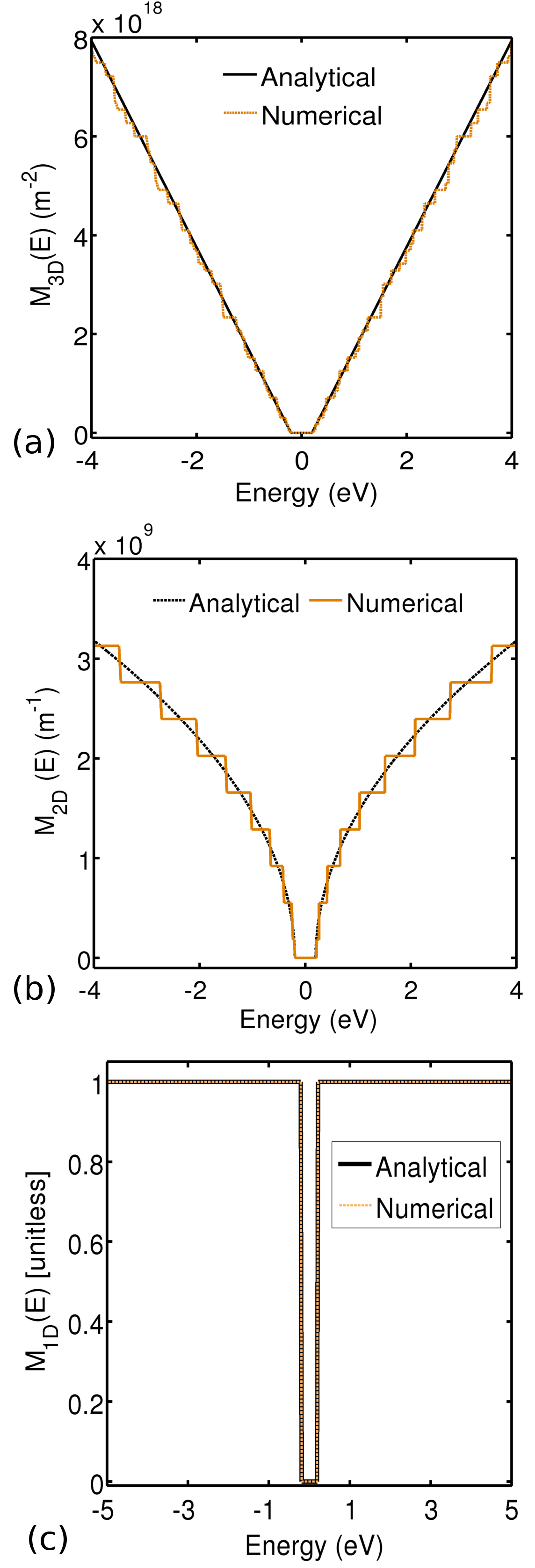}
\caption{Comparison of the numerical modes calculation using the algorithm with analytical modes calculation using parabolic bands with  $m^{*}=m_{0}$ (from Ref \cite{kim_dim_TE}) for (a) 3D, (b) 2D and (c) 1D structure. The steps in the 2D case appear due to the sparse energy grid chosen. }
\label{fig:modes_compare_anal}
\end{figure}

\begin{figure}[!hbt]
\centering
\includegraphics[width=2.1in,height=5.0in]{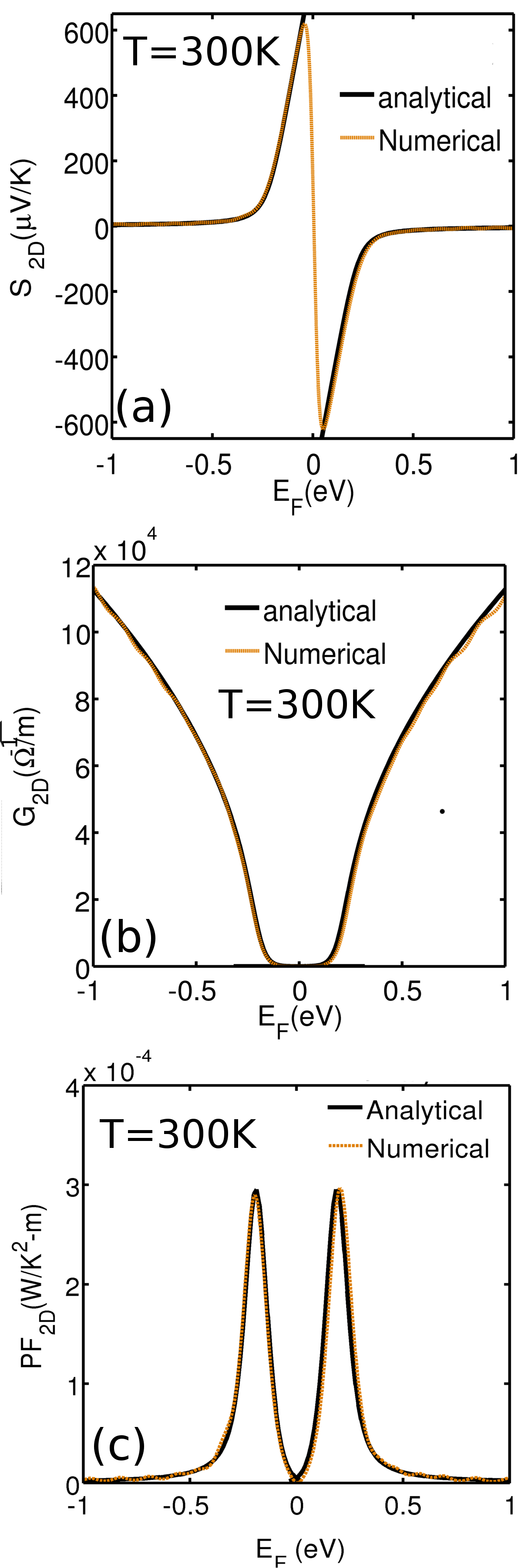}
\caption{
Comparison of the numerical calculation with analytical expression for effective mass from Ref \cite{kim_dim_TE} for a 2D system for (a) Conductance (b) Seebeck  Coefficient and (c) Thermoelectric Power Factor at 300K. The numerical results compare within 1\% to the analytical values. The parameters used for the parabolic bands are provided in Table \ref{tab_eff_param}.}
\label{fig:modes_compare_2D}
\end{figure}

The parameters used for the electronic energy dispersion are shown in Table \ref{tab_eff_param}. For all dimensions the number of energy points in the DOM (EGD) is kept constant at 500. The analytical results for the modes and TE parameters are obtained from Ref. \cite{jeong_electron} and \cite{kim_dim_TE}.  The number of modes for all three dimensions compare very well ($\le$ 4\% error) with the analytical results as shown in Fig. \ref{fig:modes_compare_anal}. Using the modes, the TE values are calculated. Only the 2D case is shown in Fig. \ref{fig:modes_compare_2D}. The agreement with the analytical calculations is very good with around 1\% error in the numerical values. The 3D and 1D cases also show very good agreement with the analytical calculations. Thus, the algorithm provides accurate results for the electronic TE parameters. 

\subsubsection*{Realistic Bands: Bulk $Bi_2Te_3$}
\begin{figure}[!htb]
\centering
\includegraphics[width=3.2in,height=1.9in]{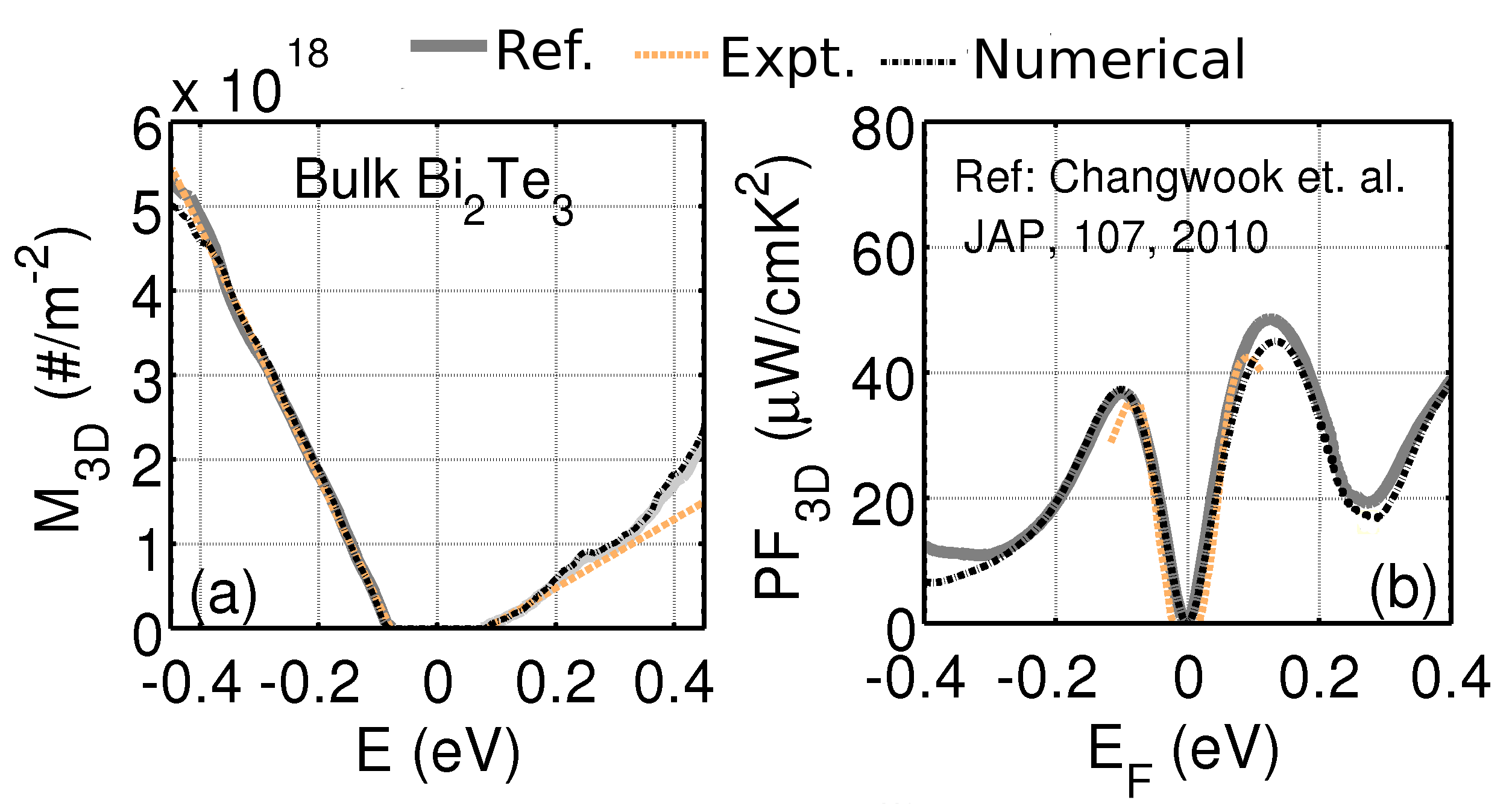}
\caption{Comparison of (a) DOM calculation and (b) Power factor at 300K for $Bi_2Te_3$, using the algorithm, with the theoretical calculations reported in Ref. \cite{jeong_electron} and experimental results from \cite{Bite_exp_electron}. The electronic energy dispersion for  bulk $Bi_2Te_3$ is obtained using the $sp^3d^5s^*$ tight-binding model \cite{BiTe_TB_1}. The PF matching for $Bi_2Te_3$ is obtained assuming a constant mean-free-path of 18, 4 nm for conduction and valence bands respectively as reported in Ref. \cite{jeong_electron}.}
\label{fig:BiTe_TE_electron}
\end{figure}

\begin{figure}[!hbt]
\centering
\includegraphics[width=3.2in,height=1.8in]{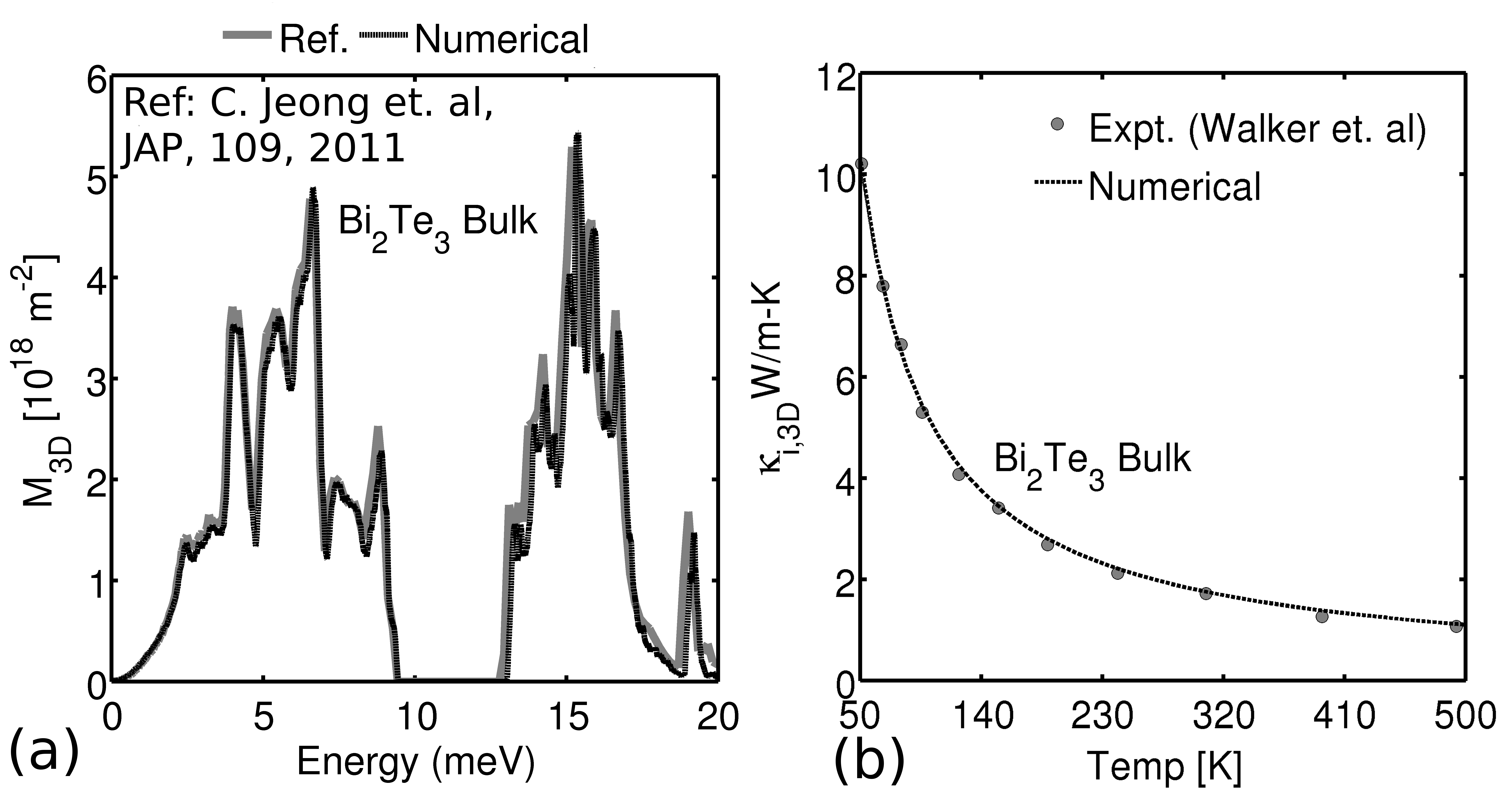}
\caption{(a) Comparison of the bulk $Bi_2Te_3$ phonon modes calculated using the algorithm and theoretical value reported in Ref. \cite{jeong_phonon}. The bulk phonon dispersion is obtained using GULP \cite{gulp}. (b) Comparison of the simulated and experimental \cite{Bite_ktherm_exp} thermal conductivity for Bulk $Bi_2Te_3$ from 50 to 500K. The phonon scattering mechanisms considered here are outlined in detail in Ref. \cite{jeong_phonon}.}
\label{fig:BiTe_TE_phonon}
\end{figure}

As a next step of verification, the algorithm is tested for the calculation of TE parameters for $Bi_{2}Te_{3}$. The same Tight-binding electronic dispersion \cite{BiTe_TB_1} is used for the calculation of the DOM as used in Ref. \cite{jeong_electron}. The agreement with the published DOM results is very good (within 1\% error) as shown in Fig. \ref{fig:BiTe_TE_electron}a. Using the DOM, the $S$ and $G$ are also calculated which are used to obtain the $PF$. The calculated $PF$ again shows a very good agreement with the published theoretical result \cite{jeong_electron} as well as with the experimental data \cite{Bite_exp_electron} as shown in Fig. \ref{fig:BiTe_TE_electron}b. The same calculation is also performed for the lattice thermal conductivity of bulk $Bi_2Te_3$. The phonon dispersion is obtained using GULP \cite{gulp} as provided in Ref. \cite{jeong_phonon}. The agreement of the calculated phonon modes with the published theoretical result \cite{jeong_phonon} is very good (Fig. \ref{fig:BiTe_TE_phonon}a). Also the lattice thermal conductivity calculated using the method provided in Ref. \cite{jeong_phonon} gives a very good agreement with the experimental value \cite{Bite_ktherm_exp}. Thus, the present algorithm provides accurate TE values for real materials too.

\section{Summary and Outlook}
\label{sec:conc}

An efficient algorithm to calculate the electron and phonon modes in any dimension is presented. The algorithm provides an efficient implementation of a TE parameters calculation scheme based on the Landauer's approach and will be extremely useful in readily and accurately evaluating the existing as well as new thermoelectric materials. 
%The code is based on Landauer’ s approach and it works for ballistic as well diffusive case. 
The algorithm is sensitive to the transverse K point density in the E(k) relation both in terms of the final TE calculations as well as the compute time. A proper optimization of the K-point reduction is provided to allow for fast and accurate TE parameter calculations. The results from the algorithm are also bench-marked with analytical and real material TE parameter values. This algorithm will be useful for developing computer programs to evaluate the TE performance of new and artificial materials in the future.

%Density of states, is also calculated, which is needed for the calculation of carrier densities and scattering time constant like ADP (~ DOS). 
%To further reduce the DOM calculation time, parallel computation of the modes for the $K_{\perp}$ directions can be adopted as proposed in Ref. \cite{ele_phon_mathieu}. 

\section*{Acknowledgments}

The authors would like to acknowledge the financial support from FCRP-MSD under Semiconductor
Research Corporation (SRC), Nano Research Initiative (NRI) under Midwest Institute for Nanoelectronics Development
(MIND) and National Science Foundation (NSF). Computational resources provided by nanoHUB.org, a portal funded by NSF
under the Network for Computational Nanotechnology (NCN), is also acknowledged. 



%% References


\bibliographystyle{IEEEtran} %elsarticle-num-names}
\bibliography{refs}

\newpage

\end{document}